\documentclass[12pt]{iopart}
\usepackage{graphicx}
\newcommand{\ecm}{\, e \, {\rm cm}}
\begin{document}

\title[Searches for EDMs]{A global perspective on searches for Electric Dipole Moments}

\author{Guillaume Pignol}

\address{Univ. Grenoble Alpes, CNRS, Grenoble INP, LPSC-IN2P3, 38000 Grenoble, France}
\ead{guillaume.pignol@lpsc.in2p3.fr}
\vspace{10pt}
\begin{indented}
\item[]September 2019
\end{indented}

\begin{abstract}
Many experiments are underway in the world to search for a non-zero electric dipole moment (EDM) of a particle with spin 1/2 such as the neutron or the electron. 
Finding an EDM would reveal new sources of CP violation. 
EDM measurements are motivated by the high sensitivity to new physics beyond the Standard Model. 
They are relevant to find the explanation for the matter-antimatter asymmetry of the Universe. 
A variety of programs with different systems are being pursued, with free neutrons, diamagnetic atoms, paramagnetic systems, and charged particles in storage rings. 
This article presents a basic introduction of the subject and attempts to compile the ongoing projects. 
\end{abstract}

\section{Introduction}

The electric dipole moment (EDM) $\vec{d}$ of a composite system measures the separation of the positive and negative electric charges, it is associated with an energy $-\vec{d}\cdot \vec{E}$ in an external electric field. 
In fact that interaction term can be taken as the definition of the EDM, even for a non-composite system such as an electron. 
For any simple system of spin 1/2, the EDM, being a vector operator, must be proportional to the Pauli matrices $\hat{\vec{\sigma}}$ acting on the spin states. 
The Hamiltonian of a spin 1/2 particle in an electric field is
\begin{equation}
\label{hamiltonian}
\hat{H} = - d \ \hat{\vec{\sigma}} \cdot \vec{E}, 
\end{equation}
where $d$ is the permanent electric dipole moment of the particle. 
Hence, the EDM of a simple particle really quantifies the coupling between the spin and an applied electric field, in the same way that the magnetic dipole moment quantifies the coupling between the spin and a magnetic field. 

\begin{figure}
\center
\includegraphics[width = .87\linewidth]{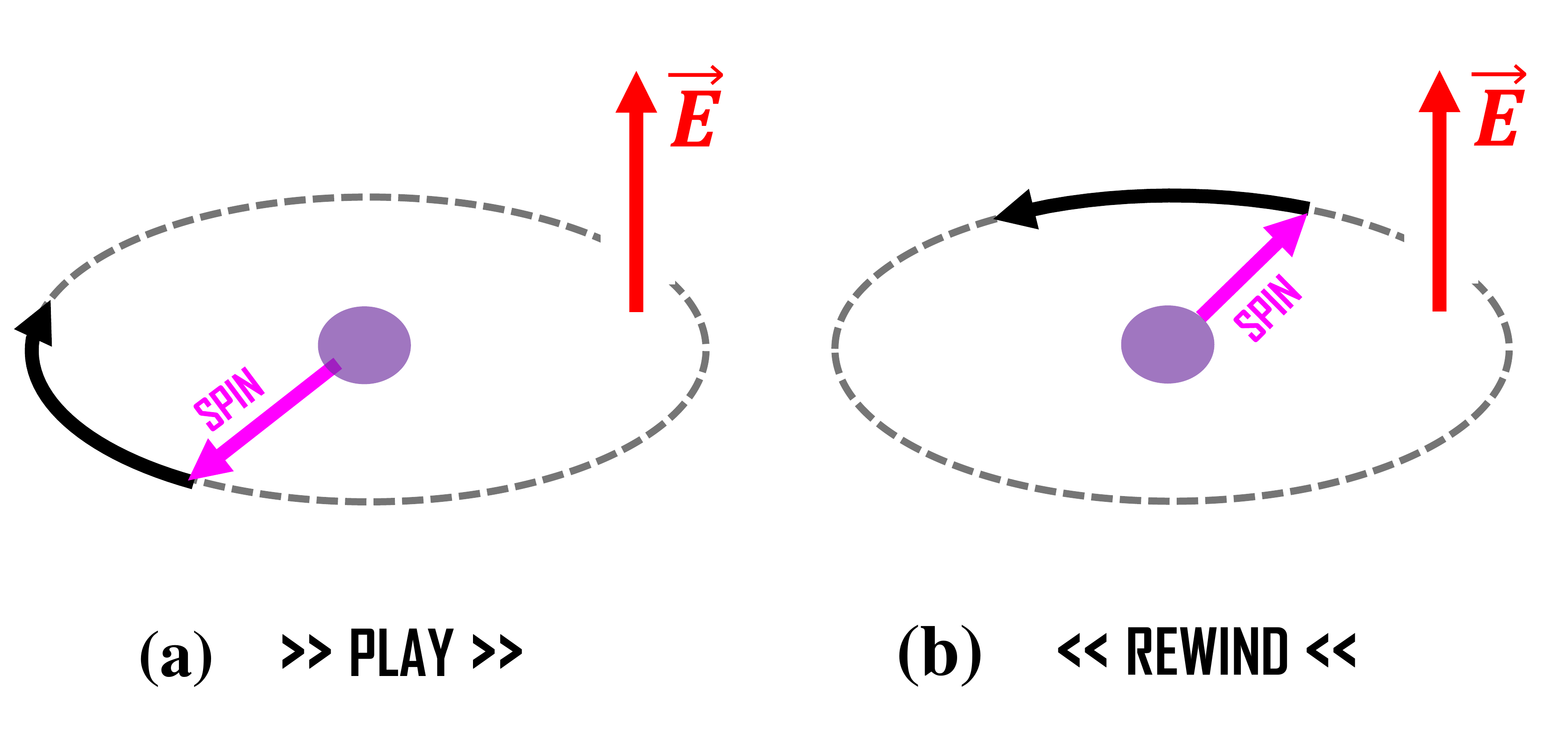}
\caption{\label{fig_1}
(a) Evolution in an electric field of a particle spin with a non-zero - positive in this case - EDM. 
(b) Time-reversed version of the evolution (a). 
The fact that (a) and (b) are different constitutes a violation of time reversal symmetry. }
\end{figure}

The coupling (\ref{hamiltonian}) results in the dynamics shown in figure \ref{fig_1} (a), for a spin  initially perpendicular to the electric field. 
The spin precesses around the field at an angular frequency given by $\hbar \omega = 2 d E$. 
As shown in figure \ref{fig_1} (b), the mere existence of a non-zero EDM would constitute a violation of time reversal symmetry, because spin precession in an electric field discerns the past and the future.  

Now, despite decades of experimental efforts, the many measurements of the EDMs of various particles are all compatible with zero. 
Permanent EDMs, if they exist, are extremely tiny. 
For example, the current limit on the magnitude of the neutron EDM is \cite{Pendlebury2015}
\begin{equation}
|d_n| < 3 \times 10^{-26} \ecm \ {\rm (90 \% \ C.L.).}
\end{equation}
In a large electric field of $10$~kV/cm, it would take more - much more? - than 80 days for the spin precession to complete one full turn. 

This paper gives a global overview of the quest for a non-zero EDM, an active field of experimental research today. 
It updates previous overviews on EDM searches \cite{Jungmann2013, Kirch2013}. 
For a more in-depth treatment of the subject, the reader should consult the recent review \cite{Chupp2019}. 
In section \ref{sec_theo} we will explain the relevance of the EDM searches in particle physics and cosmology. 
There are many experimental efforts underway worldwide to improve the sensitivity of the EDM searches using various systems. 
The archetype is the neutron EDM, that we will cover in section \ref{sec_neutron}. 
In the following sections we will cover the EDM searches with diamagnetic atoms, paramagnetic systems, and finally with charged particles. 

\section{Relevance of the EDM quest in particle physics and cosmology}
\label{sec_theo}

From the point of view of relativistic field theory, the EDM of a fermion $f$ corresponds to the following coupling to the electromagnetic field $F_{\mu \nu}$: 
\begin{equation}
\label{lagrangian}
\mathcal{L}_{\rm EDM} = - \frac{id}{2} \bar{f_L} \sigma^{\mu \nu} f_R F_{\mu \nu} + h.c.
\end{equation}
where $f_L$ and $f_R$ are the left and right chirality components of the fermion. 
In the non-relativistic limit the lagrangian density (\ref{lagrangian}) reduces to the hamiltonian (\ref{hamiltonian}). 
We note that the coupling (\ref{lagrangian}) explicitly violates CP symmetry if $d$ is non-zero. 
It is consistent with the fact that the hamiltonian (\ref{hamiltonian}) violates the time reversal symmetry and the CPT theorem which states that T-violation is equivalent to CP-violation in any local relativistic quantum field theory. 

\begin{figure}
\center
\includegraphics[width = .87\linewidth]{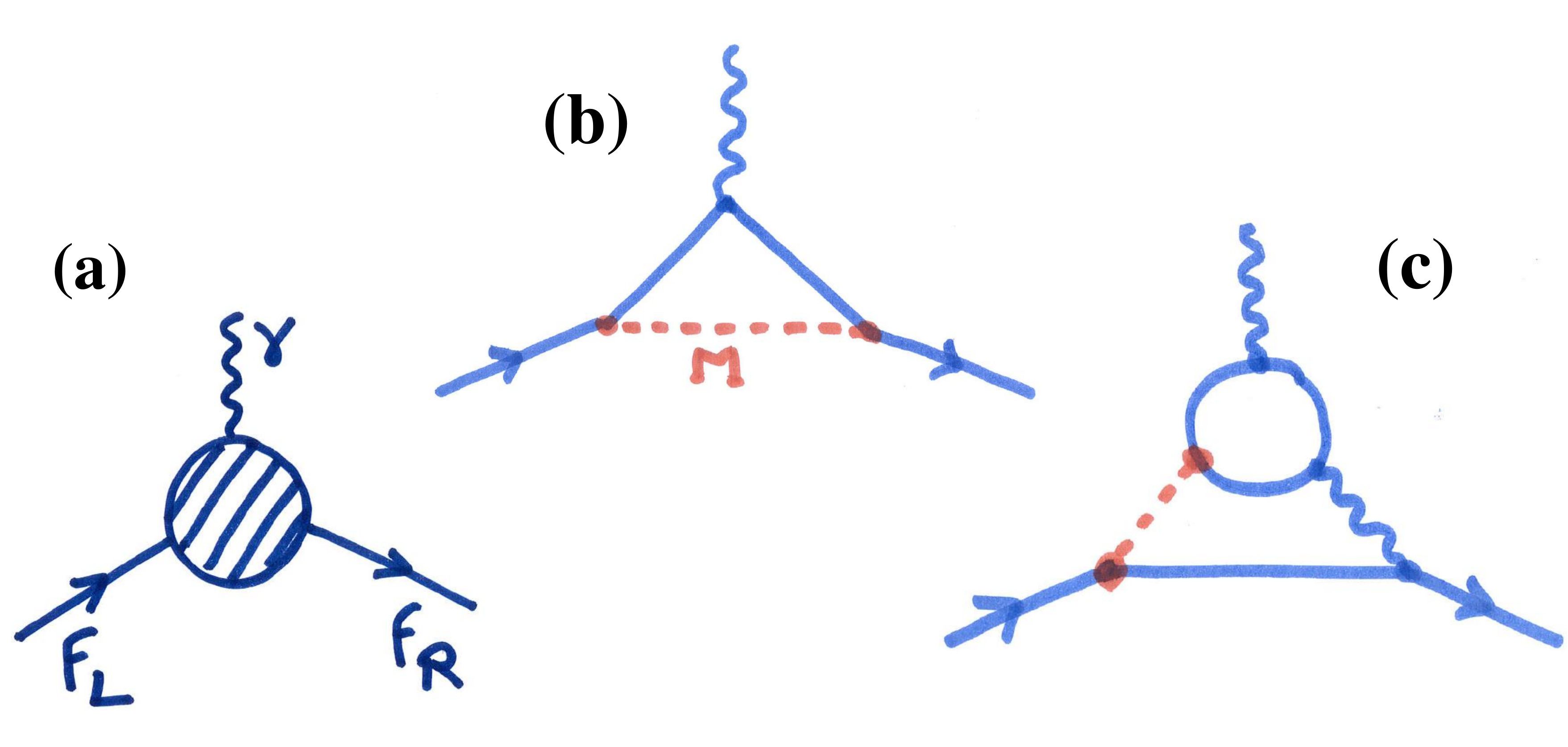}
\caption{\label{fig_2}
(a) Feynman diagram corresponding to the EDM coupling (\ref{lagrangian}). 
(b) Example of a one-loop diagram contributing to the fermion EDM. 
(c) Two-loop Barr-Zee diagram  contributing to the fermion EDM. 
}
\end{figure}

The coupling (\ref{lagrangian}), also represented in figure \ref{fig_2}~(a), is an effective  non-renormalizable interaction which could be generated by the effect of virtual particles. 
Figure \ref{fig_2}~(b) shows a possible diagram involving the virtual exchange of a heavy boson of mass $M$ and with a complex coupling $g e^{i \phi}$
 to the fermion. 
It generates an EDM of $d \approx e \hbar c \, g^2/(4 \pi)^2 \sin(\phi)\cos(\phi) \, m_f / M^2$. 
This formula can be used to estimate the order of magnitude for the EDM of the first generation fermions -- say the $d$ quark ($m_f = 5$~MeV) -- induced by a boson at the TeV scale ($M\approx1$~TeV and $g^2/(4\pi) \approx 10^{-2}$) with maximal CP violation $(\sin(\phi) \approx 1)$: we get $d\approx 10^{-25} \ecm$. 
Therefore generic CP violation above the electroweak scale is positively detectable by EDM experiments. 

The non-detection of EDMs reflects the peculiar structure of CP violation in the Standard  Model which structurally contains two sources of CP violation: a complex phase in the CKM matrix and the strong phase $\theta_{\rm QCD}$. 
EDMs induced by the CKM phase are theoretically undetectably small. 
This is due to the flavour structure of the electroweak theory: only diagrams involving all three generations of quarks in the loops can contribute to the EDM, this results in a big suppression. 
On the contrary, the strong phase induce in principle large hadronic EDMs. 
The non-observation of the neutron EDM results in the bound $|\theta_{\rm QCD}| < 10^{-10}$. 
The fact that the strong phase is measured to be unnaturally small constitutes the \emph{strong CP problem}. 
It is believed that an unknown dynamics beyond the Standard Model is at play to set this phase to zero.

\begin{figure}
\center
\includegraphics[width = .87\linewidth]{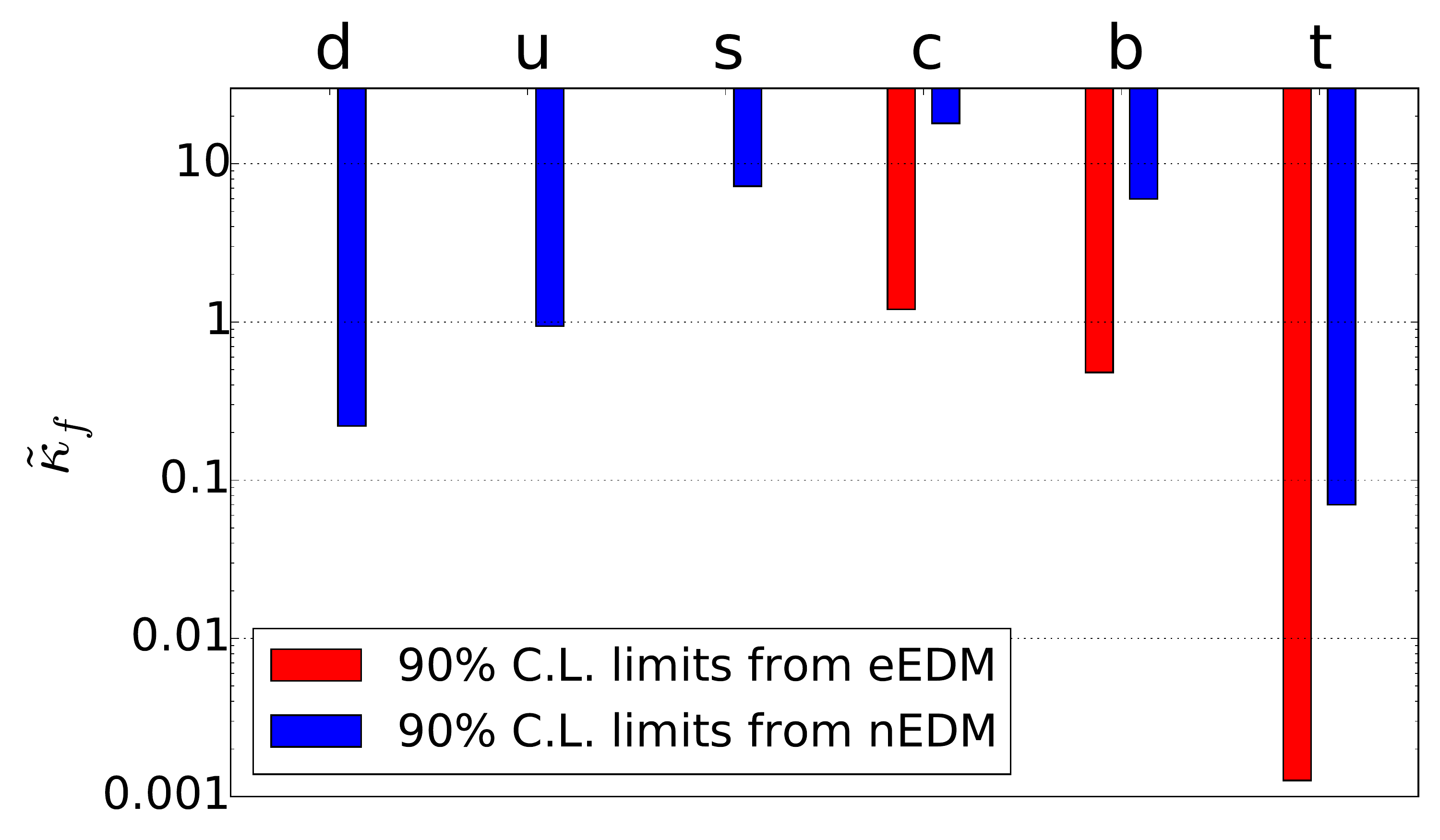}
\caption{\label{fig_3}
Current limits on the CP-violating couplings of the Higgs boson for the six quark flavours derived from the electron EDM (red bars) and from the neutron EDM (blue bars), adapted from \cite{Brod}. }
\end{figure}

EDMs are sensitive probes of CP violation effects beyond the Standard Model with practically zero background from the CKM phase. 
As a concrete example let us consider the search for CP-violating couplings of the Higgs boson $h$ to fermions. 
The Higgs couplings are generically parameterized by the following lagrangian
\begin{equation}
\mathcal{L}_h = - \frac{y_f}{\sqrt{2}} \left( \kappa_f \bar{f} f h + i \tilde{\kappa}_f \bar{f} \gamma_5 f h \right), 
\end{equation}
where $y_f$ is the Yukawa coupling of the fermion $f$, $\kappa_f$ and $\tilde{\kappa_f}$ are the CP-conserving and CP-violating coupling constants. 
The Standard Model predicts $\kappa_f = 1$ and $\tilde{\kappa}_f = 0$. 
This coupling generates EDMs though the two-loops diagram shown in figure \ref{fig_2} (c). 
The limits on the CP-violating couplings to the quarks derived from the neutron and electron EDM bounds are shown in figure \ref{fig_3}. 
This plot illustrates the complementarity of EDM searches: the electron EDM is more sensitive to $\tilde{\kappa}$ of the heavy quarks while the neutron EDM is more sensitive to $\tilde{\kappa}$ of the light quarks. 
It also illustrates the great sensitivity of EDM searches: fundamental CP-violating couplings of order unity, relative to CP-conserving couplings, are already excluded except for the $s$ quark. 
Next generations of EDM experiments will push these limits down by an order of magnitude, or perhaps discover a signal induced by small CP-violation in the Higgs sector. 

Let us complete this section by emphasizing the importance of searching for new sources of CP-violation. 
First, this is a generic feature of models extending the SM, which inevitably come with additional complex (therefore CP-violating) free parameters. 
More compellingly, cosmology actually demands new CP violation sources to solve the baryon asymmetry puzzle. 
Several classes of possible baryogenesis models have been invented to explain the generation of the matter-antimatter asymmetry in the early Universe. 
They almost all have in common to satisfy Sakharov's necessary conditions: (i) process out of thermal equilibrium, (ii) existence of baryon number violation processes, (iii) existence of C and CP violating interactions. 
An appealing possibility, called 
\emph{Electroweak baryogenesis}, poses that baryogenesis occurred at the electroweak phase transition epoch of the Universe, at a temperature of about $100$~GeV. 
See \cite{Cline2017} for a recent discussion on the subject.  
For baryogenesis to work, new CP-violating interactions must have been active at this temperature, therefore the mass of the new particles could not be much heavier than $1$~TeV and and the CP-violating interaction they mediate should be sufficiently strong. 
The models therefore also predict sizable EDMs and the future EDM experiments will either discover a nonzero EDM or exclude most of electroweak baryogenesis models. 

\section{Search for the neutron EDM}
\label{sec_neutron}

The history of EDM searches started with the neutron in the 1950's. 
The basic idea is to use polarized neutrons and measure precisely the spin precession frequency $f$ in parallel or antiparallel magnetic and electric fields: 
\begin{equation}
f = \frac{\mu}{\pi \hbar} B_0 \pm \frac{d}{\pi \hbar}E. 
\end{equation}
The EDM term can be separated from the much larger magnetic term by taking the difference of the frequency measured in parallel and antiparallel configurations. 
As we discussed in the introduction, the EDM term is very small ($d E / \pi \hbar \approx 10^{-7}$~Hz for $d = 10^{-26} \ecm$ and $E=15$~kV/cm) compared to the magnetic term (typically, $f=29$~Hz for $B_0 = 1 \, \mu$T). 
To detect such a minuscule coupling, one needs (i) a long interaction time of the neutrons with the fields, (ii) a high flux of neutrons and (iii) a precise control of the magnetic field. 
The first experiment by Smith, Purcell and Ramsey \cite{Smith1957} used a beam of thermal neutrons passing in the electric field during $T \approx 1$~ms. 
The precession time could be greatly increased by using \emph{ultracold neutrons} (UCNs). 
These are neutrons with a kinetic energy smaller than the neutron optical potential of solid materials, typically 100~neV. 
These neutrons can therefore be stored in material traps because they undergo total reflection upon collision with the walls of the trap. 
In the best previous measurement \cite{Pendlebury2015} performed at ILL in the period 1998-2002, UCNs were stored in a chamber permeated by a weak magnetic field and a strong electric field during $T \approx 100$~s. 
Although the systematic error is also a big concern, this measurement was limited by the statistical error and thus by the intensity of the ILL/PF2 UCN source. 
New higher intensity UCN sources are now coming online at several major neutron factories worldwide, which are exploited by several nEDM projects. 
In particular, the nEDM experiment has collected data \cite{Abel2018_nEDM} in 2015-2016 at the PSI UCN source, which will result in a slightly improved measurement of the neutron EDM (the analysis is still ongoing at the time of writing). 
Other ongoing nEDM projects  \cite{Ahmed2019,Abel2018_n2EDM,Ito2017,Ahmed2018,Serebrov2017,Chanel2018} are listed in table \ref{EDM_projects}, they are all at a different stage of readiness and they aim at an improvement in sensitivity by a factor 10 to 100 compared to the previous measurement \cite{Pendlebury2015}. 
More details on nEDM searches can be found in the recent reviews \cite{Schmidt-Wellenburg2016,Filippone2018}.

\begin{table}
\caption{\label{EDM_projects}List of active ongoing projects \cite{EDMworld} searching for the EDM of the neutron, diamagnetic atoms (Hg, Xe, Ra),  paramagnetic systems and charged particles. }
\begin{tabular}{@{}llll}
\br
project & location & concept &  references \\
\mr
nEDM@SNS  & Oak Ridge spallation source & UCN in superfluid helium & \cite{Ahmed2019} \\
n2EDM     & PSI spallation source & UCN  double chamber & \cite{Abel2018_n2EDM} \\
nEDM@LANL & Los Alamos spallation source & UCN double chamber & \cite{Ito2017} \\
panEDM    & ILL reactor Grenoble & UCN  double chamber &  \cite{Wurm2019} \\
TUCAN     & TRIUMF spallation source & UCN double chamber & \cite{Ahmed2018} \\
PNPI nEDM      & ILL - PNPI   & UCN double chamber & \cite{Serebrov2017} \\
beam nEDM & ESS spallation source & pulsed cold neutron beam & \cite{Chanel2018}\\
\hline
Hg EDM & Seattle & vapor cells mercury-199 & \cite{Graner2016} \\
quMercury & Bonn & laser cooled mercury-199 & \\
MIXed & J\"ulich - Heidelberg &  xenon-129 + helium-3 & \cite{Allmendinger2019} \\
HeXeEDM & Berlin &  xenon-129 + helium-3 & \cite{Sachdeva2019} \\
Xe EDM & Riken & xenon-129 + xenon-131 & \cite{Sato2018} \\
Ra EDM & Argonne & laser-cooled radium-225 & \cite{Bishof2016} \\
\hline
Cs,Ru EDM & Penn State & trapped cold alkali & \cite{Tang2018} \\ 
Fr EDM & CYRIC, Riken & laser-cooled francium & \\
EDM$^3$ & Toronto & BaF within a rare gas matrix & \cite{Vutha2018} \\
NL-eEDM & Nikhef & BaF cold beam & \cite{Aggarwal2018} \\
JILA EDM & Boulder & trapped molecular ions HfF$^+$ ThF$^+$ & \cite{Cairncross2017} \\
ACME & Yale &  cryogenic ThO beam & \cite{Andreev2018} \\
eEDM & London &  slow YbF beam &  \cite{Hudson2011} \\
\hline
CPEDM &  & proton or deuteron storage ring &  \\
muEDM & PSI & compact muon ring, frozen spin & \cite{Crivellin2018} \\
$\mu$ g-2/EDM & JPARC & compact muon ring & \cite{Abe2019} \\
$\mu$ g-2 & Fermilab & magic momentum muon ring & \cite{Chislett2016} \\
\br
\end{tabular}
\end{table}

\section{Search for the EDM of diamagnetic atoms}

Diamagnetic atoms are atoms with no net electronic spin.
Due to this property, very high precision can be obtained for the EDM of diamagnetic atoms with nuclear spin 1/2, in particular mercury-199, xenon-129 and radium-225. 

In the case of mercury-199, super-precise monitoring of the spin precession can be achieved by making use of atom-light interaction. 
The current best limit is \cite{Graner2016}:
\begin{equation}
|d_{\rm Hg}| < 7.4 \times 10^{-30} \ecm \ {\rm  (95 \% \ C.L.)}
\end{equation}

In the case of xenon-129, the measurement profit from the very long coherence time (many hours) of the spin precession. 
The current best limit is \cite{Allmendinger2019}:
\begin{equation}
|d_{\rm Xe}| < 1.5 \times 10^{-27} \ecm \ {\rm  (95 \% \ C.L.)}
\end{equation}

It is important to note that these limits apply to the atomic EDM and not the nuclear EDM. 
As stated by Schiff's theorem, a nuclear EDM is shielded by the electrons and does not generate an atomic EDM. 
Instead, atomic EDMs could possibly be generated by two sources: (i) T-violating electron-nucleon interactions, or (ii) a nonzero nuclear Schiff moment. 
The Schiff moment is a T-odd nuclear deformation which generates an electric field inside the nucleus along the spin. That electric field is pulling the electrons in $s$ orbitals therefore generating an atomic EDM. 
The Schiff moment itself could be generated by either T-violating nucleon-nucleon interactions or by a  nucleon EDM. 
Overall the effect is larger in heavy nuclei, hence the experimental focus on mercury-199 and xenon-129. 
At the end, the shielding effects are compensated by the better absolute sensitivity of experiments with diamagnetic atoms, as compared to the neutron. 
All diamagnetic systems (neutron, mercury and xenon) have comparable and complementary sensitivity to fundamental sources of CP-violation \cite{Chupp2019}. 
Also, Schiff moments are enhanced in octupole-deformed nuclei. 
This has motivated recently the search for EDMs of radioactive nuclei such as radium-225. 
The current best limit is \cite{Bishof2016}:
\begin{equation}
|d_{\rm Ra}| < 1.4 \times 10^{-23} \ecm \ {\rm  (95 \% \ C.L.). }
\end{equation}

The search for EDMs of diamagnetic atoms is very active today, with prospects to improve the sensitivity by a factor 100 in all three systems. The ongoing projects \cite{Graner2016,Allmendinger2019,Sachdeva2019,Sato2018,Bishof2016} are listed in table \ref{EDM_projects}. 

\section{EDM searches with paramagnetic atoms and polar molecules}

Paramagnetic systems, i.e. atoms or molecules with an unpaired electron, can be sensitive to  T-violating electron-nucleon interactions and to the electron EDM. 
The most sensitive probes are atoms with large $Z$, in particular cesium of radioactive francium, and heavy polar molecules like BaF, ThO, YbF, or even molecular ions HfF$^+$, ThF$^+$. 
We refer to the recent review \cite{Safronova2017} for more details about the search for EDMs with atoms and molecules. 
The current best limit on the electron comes from the ACME experiment with ThO molecule \cite{Andreev2018}: 
\begin{equation}
|d_e| < 1.1 \times 10^{-29} \ecm \ {\rm  (90 \% \ C.L.). }
\end{equation}
There are several projects \cite{Tang2018,Vutha2018,Aggarwal2018,Cairncross2017,Andreev2018,Hudson2011}, listed in table \ref{EDM_projects}, aiming at improving the sensitivity on the electron EDM by a factor of 100 or more.

\section{Search for the EDM of charged particles in a storage ring}

Polarized charged particles, in particular protons, deuterons or muons, can be confined in circular storage rings with either a radial electric field, or a vertical magnetic field, or a combination of both. 
It is apparently not a good situation to measure an EDM since the electric and magnetic fields cannot be made parallel or antiparallel and the classic EDM search strategy does not work for charged particles. 
The situation is in fact more complicated, because contrary to classic EDM searches with particles practically at rest, the relativistic motional fields $\vec{E} \times \vec{v}$ and $\vec{B} \times \vec{v}$ are not small for charged particles in a storage ring. 
In the frame rotating with the cyclotron motion, the precession vector of the spin is given by the BMT equation: 
\begin{equation}
\vec{\omega} = \frac{q}{m} \left[ a \vec{B} - \left( a + \frac{1}{1-\gamma^2}\right) \vec{v} \times \vec{E} \right] + 2d \left[ \vec{v} \times \vec{B} + \vec{E} \right], 
\end{equation}
where $q$ is the charge of the particle, $m$ its mass and $a = (g-2)/2$ is the magnetic anomaly. 
The first term is due to the magnetic dipole. 
The second term is due to the electric dipole, it makes the spin move out of the plane of the ring. 
The EDM signal corresponds to a build up of the vertical component of the spin. 
In some cases it is possible to enhance the sensitivity of the search by setting the first magnetic term to zero, a technique called the \emph{frozen spin}, by an appropriate choice of the parameters $B, E, v$. 
Ongoing projects pursuing the developments of EDM measurements with charged particles are listed in table \ref{EDM_projects}. 

\section{Conclusion}

\begin{figure}
\center
\includegraphics[width = 1.\linewidth]{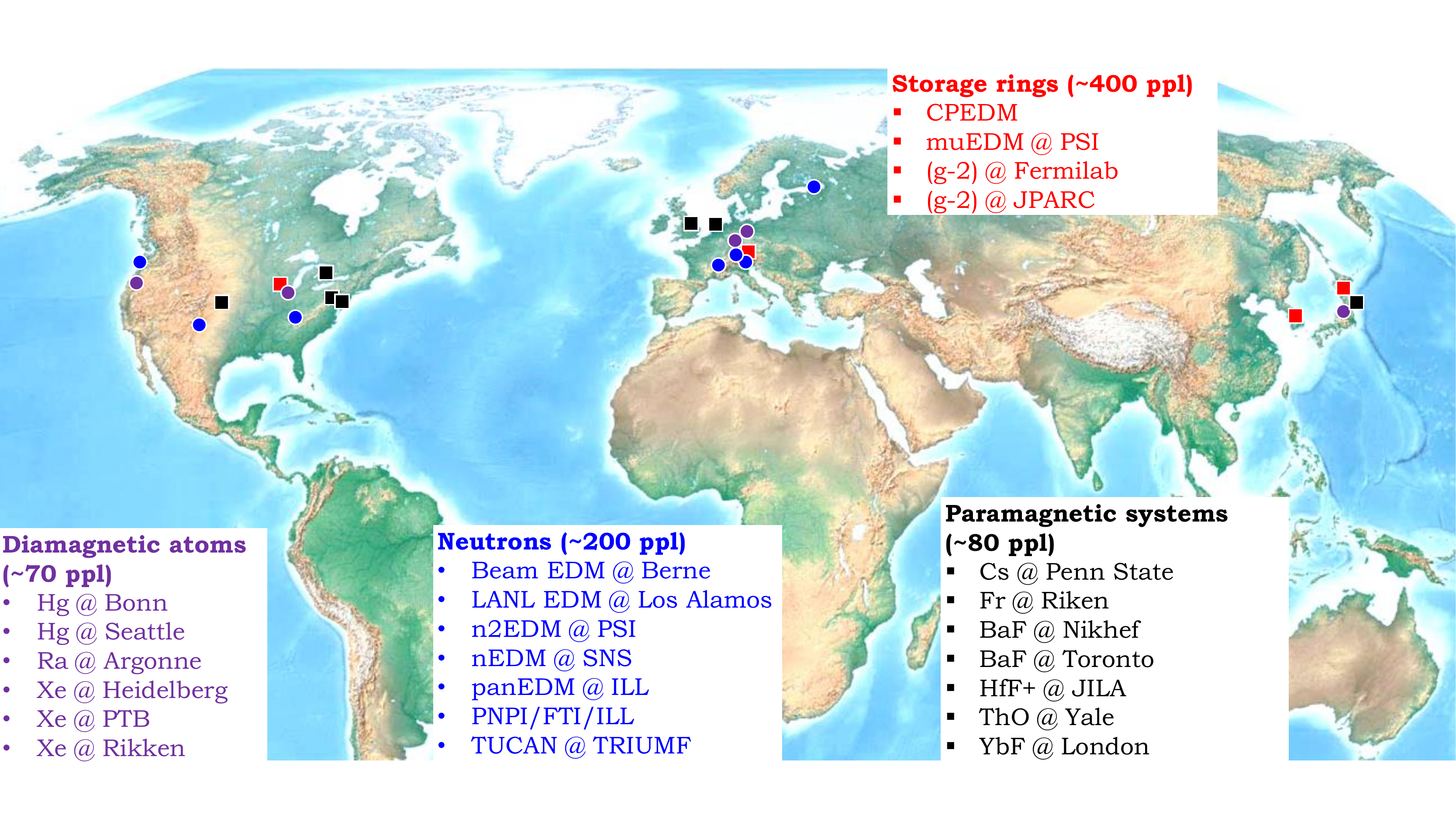}
\caption{\label{fig_4}
The world view on EDM searches, with an estimate of the number of physicists involved, adapted from \cite{EDMworld}. }
\end{figure}

The search for a non-zero fundamental electric dipole moment is an interdisciplinary field.  
The motivation comes from particle physics and cosmology. 
A broad range of experimental techniques are developed ranging from the large scale neutron facilities to advanced atomic physics. 
We have presented an overview of the theoretical motivations and the experimental programs to search for the EDMs with free neutrons, diamagnetic atoms, paramagnetic systems and charged particles. 
We must admit we have omitted the proposals to measure EDMs of heavy unstable particles (lepton $\tau$,  hyperons and charmed baryons) at particle colliders, due to the temporary incompetence of the author on this connected field. 
In figure \ref{fig_4} we show the world map of the ongoing EDM projects, with an estimate of the number of scientists involved. 
The diversity of the experiments promises exciting prospects for the future, and maybe a discovery of fundamental importance. 

\ack
I am grateful to Dieter Ries for the compilation of the ongoing EDM projects used to fill table \ref{EDM_projects}. 
I wish to thank J\'er\'emie Quevillon for his reading of the theoretical part. 
This work is supported by the European Research Council, ERC project 716651 - NEDM.

\end{document}